\documentclass{iau}
\usepackage{graphicx}
\usepackage{booktabs}

\title[Improving distances to nearby bright stars] 
{Improving distances to nearby bright stars: Combining astrometric data from Hipparcos, Nano-JASMINE and Gaia}

\author[D. Michalik, L. Lindegren, D. Hobbs, U. Lammers and Y. Yamada]   
{Daniel Michalik$^1$,
 Lennart Lindegren$^1$,
 David Hobbs$^1$,\\
 Uwe Lammers$^2$,
 \and Yoshiyuki Yamada$^3$}

\affiliation{$^1$Lund Observatory, Lund University, Box 43, SE-22100 Lund, Sweden\\e-mail: {\tt daniel.michalik, lennart, david@astro.lu.se} \\[\affilskip]
$^2$European Space Agency (ESA/ESAC), P.O. Box 78, ES-28691 Villanueva de la Ca\~{n}ada, Madrid, Spain, e-mail: {\tt uwe.lammers@sciops.esa.int} \\[\affilskip]
$^3$Department of Physics, Kyoto University, Oiwake-cho Kita-Shirakaw Sakyo-ku, Kyoto, 606-8502 Japan, e-mail: {\tt yamada@amesh.org}}

\pubyear{2012}
\volume{289}  
\pagerange{???--???}
\jname{Advancing the physics of cosmic distances}
\editors{Richard de Grijs \& Giuseppe Bono, eds.}
\begin{document}

\maketitle

\begin{abstract}
Starting in 2013, Gaia will deliver highly accurate astrometric data,
which eventually will supersede most other stellar catalogues in
accuracy and completeness. It is, however, limited to observations from
magnitude 6 to 20 and will therefore not include the brightest
stars. Nano-JASMINE, an ultrasmall Japanese astrometry satellite,
will observe these bright stars, but with much lower accuracy. Hence,
the Hipparcos catalogue from 1997 will likely remain the main source
of accurate distances to bright nearby stars. We are investigating
how this might be improved by optimally combining data from all three
missions in a joint astrometric solution. This would take advantage of
the unique features of each mission: the historic bright-star
measurements of Hipparcos, the updated bright-star observations of
Nano-JASMINE, and the very accurate reference frame of Gaia. The long
temporal baseline between the missions provides additional benefits
for the determination of proper motions and binary detection, which
indirectly improve the parallax determination further. We present a
quantitative analysis of the expected gains based on simulated data
for all three missions.         
\keywords{astrometry, catalogs, methods: data analysis, methods: statistical, reference systems}
\end{abstract}

\firstsection 
\section{Introduction}
The distance to a star can most directly be deduced from its trigonometric
parallax. From the ground this was only done for a few thousand very close-by stars,
but with the advent of space astrometry this picture changed
dramatically. Hipparcos (1989--1993) was the first satellite to
determine astrometric parameters (stellar positions, parallaxes and proper
motions) from space and yielded the distances to approximately
$21\,000$ stars with an uncertainty of better than 10\% (\cite{ESA1997}).
Now, 25 years later, Gaia will improve our knowledge of stellar astrometry 
significantly and provide millions of stellar parallaxes with unprecedented
accuracy. Gaia is an ESA cornerstone mission which will be launched for its nominal five
year mission at the end of 2013. It will continuously scan the sky in a
well-chosen pattern and observe up to a billion stars down to
magnitude 20. However, because of CCD saturation it will not
observe stars brighter than about magnitude 6. Hence the brightest $\sim$5000
stars are not observed by Gaia, and for these Hipparcos will continue to be a
main source of distance information. 

In addition to these two astrometry missions, the third upcoming mission is
the ultra-small Japanese satellite Nano-JASMINE. Just like Gaia, this satellite is
based on CCD detections and its scanning principle and observing strategy are
derived from the Gaia pendants. This mission, however, is meant to be a
technology demonstrator for larger follow-up missions and is therefore significantly
smaller and less accurate. It is scheduled for launch by the end of 2013 and
expected to provide astrometry for approximately one million stars in the visual magnitude range from $\sim$1
to 10 with an accuracy of $\sim$3 mas for objects of magnitude
7.5. Gaia data will be reduced using the Astrometric Global Iterative
Solution (AGIS), developed by ESA and Lund Observatory (\cite{Lindegren_etal2012}). Thanks to a collaboration between
the Nano-JASMINE Science Team and parts of the AGIS team, it is possible to
use AGIS also for the core data reduction of Nano-JASMINE.

\section{Catalogue combination by joint solution}

While Nano-JASMINE's uncertainties are not better than the uncertainties in the
Hipparcos results, significant improvements for bright star astrometry can be
made by combining the results of Hipparcos and Nano-JASMINE, thanks to the long
time baseline between the missions. 
The combination is done by incorporating the Hipparcos information directly in the astrometric solution for the Nano-JASMINE data.
This is done using the Hipparcos data and the inverse of its covariances as
starting values when accumulating the normal equations for the astrometric solution.
In contrast to {\em a posterior} catalogue combination, this ``joint solution'' combines the datasets in a statistically
optimal way (\cite{Michalik_etal2012}), 
taking into account the correlations between the different astrometric parameters. 

\begin{figure}
\centering
\includegraphics[width=0.7\textwidth]{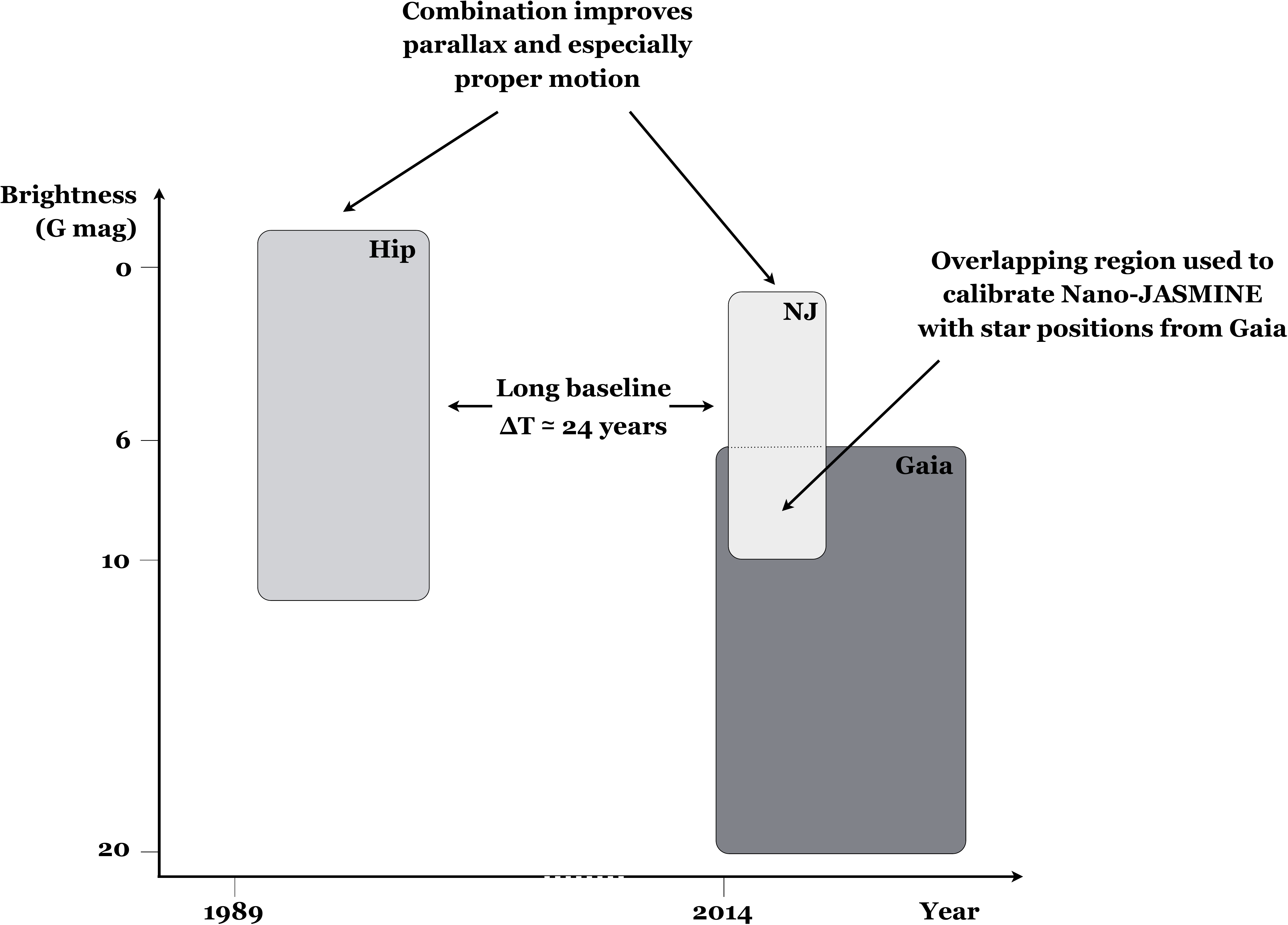}
\caption{Magnitude range versus mission time for the three astrometry missions Hipparcos (Hip), Nano-JASMINE (NJ), and Gaia.\label{michalik_fig:threecatalogues}}
\end{figure}
In addition to the astrometric improvement facilitated by the combination
described above, further improvement can be gained by incorporating (preliminary)
results of Gaia during the Nano-JASMINE data processing: Gaia and Nano-JASMINE
will both observe stars between magnitude 6 and 10. Since the astrometry of
these stars is well-determined by Gaia, a joint solution can be used to
determine the attitude and
geometry deviations of Nano-JASMINE with better accuracy and therefore improve
all Nano-JASMINE results.  This includes calibrating the basic angle (the
nominally fixed angle between the two fields of view of the satellite) which
may be affected by thermal variations originating in the low-earth-orbit of
the satellite. The basic angle stability is particularly critical to avoid
zero-point errors thus allowing the determination of absolute parallaxes.
Additionally, the Nano-JASMINE results are aligned with the Gaia reference
frame. Therefore, optimal results for bright stars are obtained by
reducing the Nano-JASMINE data together with preliminary Gaia results and by
combination of these with the historic Hipparcos measurements during data analysis (see
Fig.~\ref{michalik_fig:threecatalogues}).  We quantify the expected improvements in
parallax and proper motion determination by simulating this
scenario.

\section{Simulations\label{michalik_sec:simulations}}
\begin{figure}
\centering
\raisebox{0.35cm}{\includegraphics[width=0.34\textwidth]{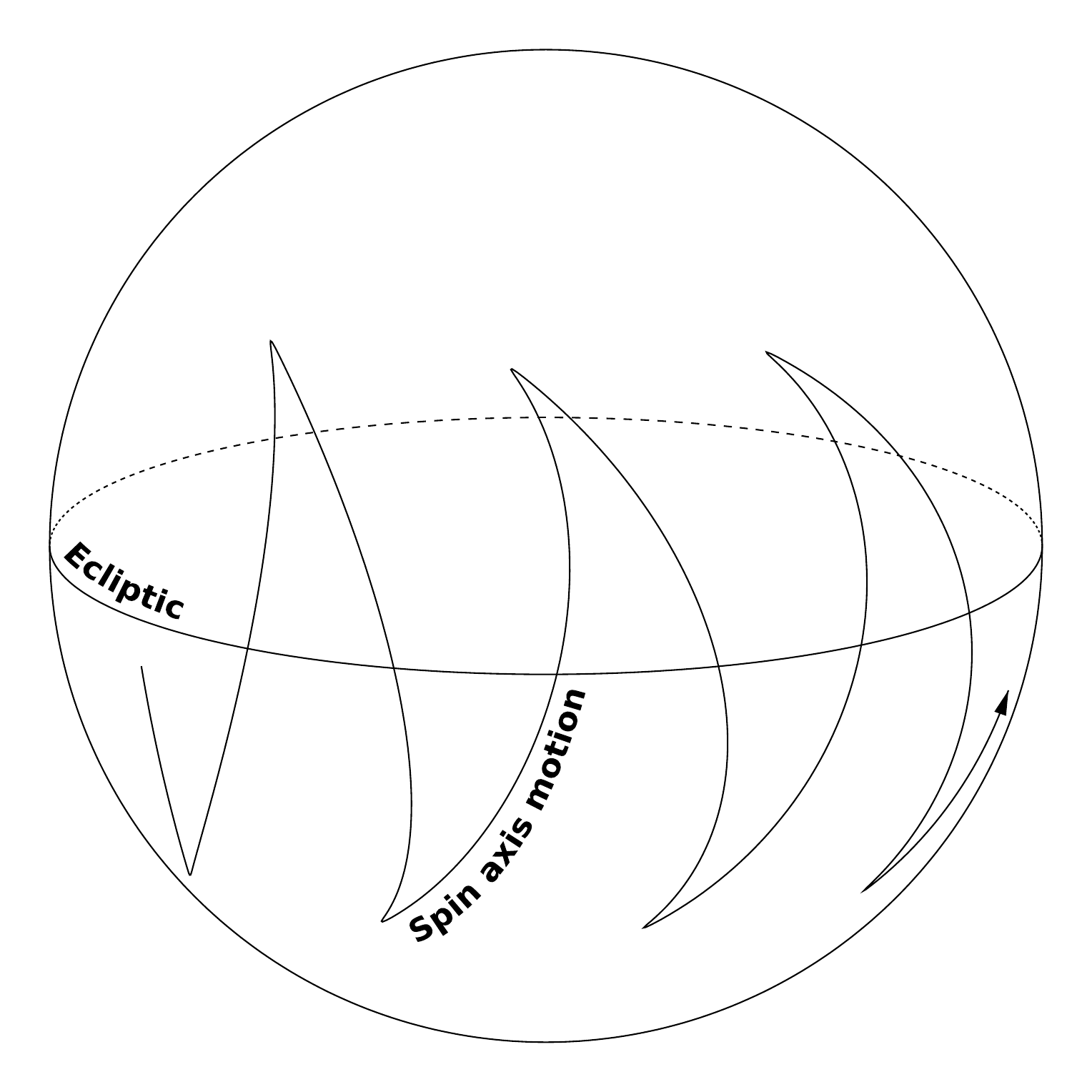}}
\includegraphics[width=0.64\textwidth]{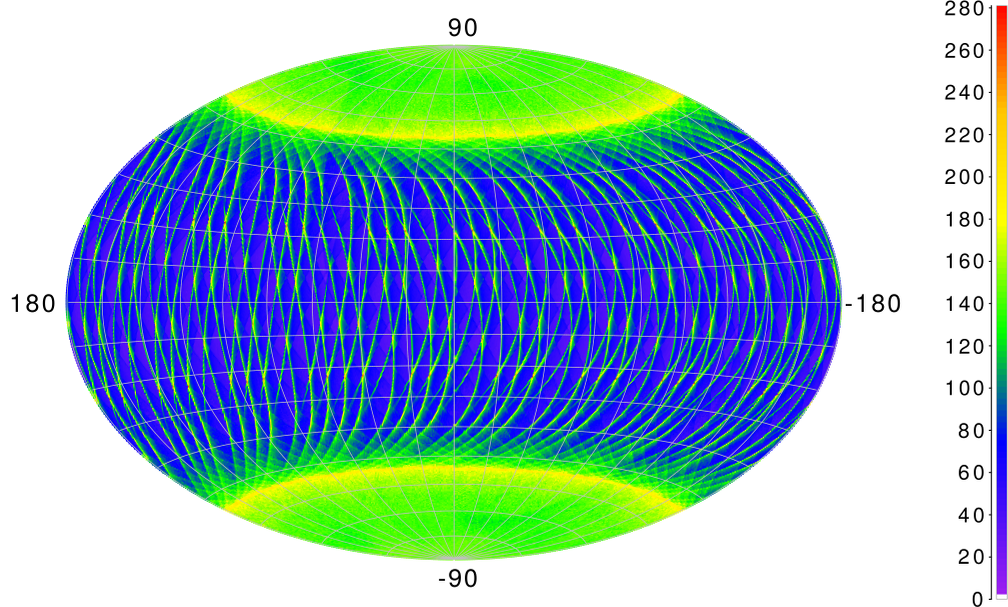}
\caption{(left) 130 days of the spin-axis motion. (right) number of observations per star (ecliptical skymap), based on simulations of two years of the baseline Nano-JASMINE scanning law.\label{michalik_fig:scanninglaw}}
\end{figure}

Simulations are carried out using AGISLab, a software package developed at Lund
Observatory to aid the development of algorithms for Gaia data processing (\cite{2012A&A...543A..15H}).
It is used to simulate Nano-JASMINE observations and to process these data in
the same manner as done for the real mission, i.e., by employing the AGIS algorithms. AGIS
is an iterative scheme that implements a block-wise least-squares solution on very
large datasets, i.e., as required to reduce data from large astrometry missions
such as Gaia. Nano-JASMINE simulations are based on a realistic scanning law
featuring a triangular spin axis precession motion (see
Fig.~\ref{michalik_fig:scanninglaw}). The observation accuracy model assumes a
(somewhat optimistic) centroiding uncertainty of 1/300th of a pixel
($\sim$7 mas) for stars brighter than magnitude 7, with additional photon noise
for fainter stars ($\sim$30 mas at magnitude 10).

The simulation dataset consists of two parts. The first is composed of the 5026 brightest Hipparcos stars (mag $<$ 6), with
astrometric parameters and uncertainties taken from the Hipparcos
Catalogue. They are used to evaluate the improvement in the uncertainties of the astrometric parameters of the bright stars. To these, we add 330\,000 randomly distributed stars of magnitude 10
that represent stars observed in the overlapping magnitude range which were not
included in the Hipparcos Catalogue. They represent the stars
that will be seen by Gaia as well as by Nano-JASMINE, and for which the
well-determined Gaia positions help to better constrain the solution.

First we simulate the errors in the Hipparcos Catalogue, providing a reference
case for the bright stars and initial values for the solutions (Cases~B and C
below), which incorporate the Hipparcos data. This is followed
by three simulations.

\begin{enumerate}
\item[Case A:] We simulate Nano-JASMINE observations with the two datasets
mentioned above and process the data without additional information, i.e.
Nano-JASMINE only.
\item[Case B:] In a second run, we incorporate the information from Hipparcos into the
data processing of Nano-JASMINE using the method described in \cite[Michalik \etal (2012)]{Michalik_etal2012}. 
\item[Case C:] In a third run we simulate the astrometric parameters with Gaia accuracy
and fold expected Gaia covariances (\cite{deBruijne2012}) into
the Nano-JASMINE data processing. This is done by setting up the covariances of
these sources with a $\sigma$ according to the expected Gaia uncertainties and
using it in a joint solution scheme. Under the optimistic assumption of uncorrelated
parameters, the covariance is set up with zeros in all
off-diagonal positions.
\end{enumerate}
\section{Results}
\begin{table}
\begin{center}
\caption{Average uncertainties (RSE\protect\footnotemark) of 5026 Hipparcos stars between magnitude 1 and 6.\label{michalik_results}}

\begin{tabular}{@{}llcc} 
\addlinespace
	\hline
&&Parallax [$\mu$as]& Proper motion [$\mu$as~yr$^{-1}$]\\
		\hline
	Reference:& Hipparcos only (Hip) & 740 & 673\\
	Case A: &Nano-JASMINE only (NJ) & 1282 & 1844\\
	Case B: &Hip + NJ & 595 & 50\\
	Case C: & \textbf{Hip + NJ + Gaia} & \textbf{588} & \textbf{43} \\
	\hline
\end{tabular}
\end{center}
\end{table}
\footnotetext{``The Robust Scatter Estimate (RSE) is
defined as 0.390152 times the difference between the 90th and 10th percentiles
of the distribution of the variable. For a Gaussian distribution it equals the
standard deviation. The RSE is used as a standardized, robust measure of
dispersion [within the Gaia core processing unit].''(\cite{Lindegren_etal2012})}

We compare the three cases and our current knowledge of the bright star astrometric uncertainties, with the
results in Table \ref{michalik_results}. Our current knowledge is
represented by the Hipparcos Catalogue. The results of the Nano-JASMINE
observations alone are less precise, but combined with Hipparcos they lead to
a significant improvement with respect to our current knowledge. Including provisional Gaia
results yields a further improvement thanks to the improved calibration of the
geometry and attitude of Nano-JASMINE. This improvement is fairly small, but it
needs to be emphasized that there is a second and very important advantage of
the Nano-JASMINE joint solution with Gaia results, i.e., alignment of the
Nano-JASMINE results with the Gaia reference frame. 

\section{Conclusions}
Nano-JASMINE offers an opportunity to significantly improve the Hipparcos
parallaxes and proper motions of the brightest $\sim$5000 stars which will not be
observed by Gaia. In addition, a combined solution with Gaia data ensures that
the results are in the same reference frame as the Gaia catalogue and that the
parallaxes are absolute. 

\end{document}